\documentclass[aps,prc,twocolumn,floatfix,superscriptaddress]{revtex4-1}
\usepackage{graphicx}
\usepackage{amsmath, enumerate}
\usepackage[pdftex,colorlinks,citecolor=blue,bookmarks]{hyperref}
\usepackage{bm}
\begin{document}

\title{Systematic study of infrared energy corrections in truncated oscillator spaces}

\author{Alexander Arzhanov} 
\affiliation{Institut f\"ur Kernphysik, Technische Universit\"at
  Darmstadt, D-64289 Darmstadt, Germany}
\affiliation{GSI Helmholtzzentrum f\"ur Schwerionenforschung,
  Planckstra{\ss}e~1, 64291 Darmstadt, Germany} 
\author{Tom\'as R. Rodr\'iguez}
\affiliation{Departamento de F\'isica Te\'orica, Universidad
  Aut\'onoma de Madrid, E-28049 Madrid, Spain}
\author{Gabriel Mart\'inez-Pinedo}
\affiliation{Institut f\"ur Kernphysik, Technische Universit\"at
  Darmstadt, D-64289 Darmstadt, Germany}  
\affiliation{GSI Helmholtzzentrum f\"ur Schwerionenforschung,
  Planckstra{\ss}e~1, 64291 Darmstadt, Germany}
\date{\today} 

\begin{abstract}
  We study the convergence properties of nuclear binding energies
  and two-neutron separation energies obtained with self-consistent mean-field calculations based on the
  Hartree-Fock-Bogolyubov (HFB) method with Gogny-type
  effective interactions. Owing to lack of convergence in a truncated working basis, we employ and benchmark one of the recently proposed infrared energy correction techniques to extrapolate our results to the limit of an infinite model space. We also discuss its applicability to global calculations of nuclear masses.
\end{abstract}
\maketitle

%
 \section{Introduction}
 \label{sec:INTRO}
%

Nuclear physics properties like nuclear masses, decay and capture
rates, fission, etc., constitute a key ingredient to study the
formation of elements in stellar nucleosynthesis. For example, all
current rapid neutron capture nucleosynthesis process
(\textit{r}-process) models require nuclear physics input for a large
number of nuclei that have extreme neutron excess and stretch up to
the limits of the nuclear chart. Such nuclei lie far beyond the
capabilities of the experimental facilities in any foreseeable future,
and hence performing \textit{r}-process simulations one has to almost
entirely rely on theoretical predictions. Since masses determine
thresholds of all nuclear reactions, the calculated final
\textit{r}-process elemental abundances of any astrophysical model are
very
sensitive~\cite{2011PhRvC..83d5809A,2016PrPNP..86...86M,2015PhRvC..92e5805M}
to the employed nuclear mass
tables. 

Self-consistent mean-field theories based on the
Hartree-Fock-Bogolyubov (HFB) variational approach with energy density
functionals (EDF) were actively developing in the recent decades and
have proven successful in the systematic study of low energy nuclear
structure~\cite{2003RvMP...75..121B,2006NuPhA.777..623P,2003RvMP...75.1021L}. In particular, the
recent HFB-based mass models~\cite{2013PhRvC..88b4308G,2009PhRvL.102x2501G} are now
found to be on a similar accuracy level in describing experimental
masses as the more phenomenological
approaches~\cite{1995ADNDT..59..185M,Moeller.Sierk.ea:2016,2011PhRvC..84a4333L}. Nonetheless,
in order to further increase the predictive power of HFB-based models,
a particular attention must be paid to the following important issues
inherent to all currently used HFB-models with either Skyrme, Gogny, or
Relativistic EDFs. First of all, there are missing correlations at the
purely mean-field level, and one has to introduce the so-called
Beyond-Mean-Field (BMF) corrections, such as symmetry restorations and/or
configuration
mixing~\cite{2009PhRvL.102x2501G,2006PhRvC..73c4322B,2008PhRvC..78e4312B,2015PhRvC..91d4315R},
in order to achieve a better compliance with experimental
data. Furthermore, nuclei with an odd number of neutrons and/or
protons are usually not treated at the same self-consistent level as
the even-even isotopes~\cite{PhysRevC.78.014304,PhysRevC.86.064313,2014PhRvL.113p2501B}, resulting in
elevated uncertainties when describing such odd-mass
nuclei. This aspect affects theoretical predictions of
  reaction $Q$-values needed to describe nucleosynthesis processes.
Finally, there are also purely numerical problems, as an incomplete
convergence of observables in practical calculations that can lead to
numerical noise in the form of artificial jumps in the calculated
binding and neutron-separation energies~\cite{2011PhRvC..83d5809A,PhysRevLett.102.152503,2015PhRvC..91d4315R}. In what follows, we discuss the issue of insufficient convergence of practical HFB
calculations in more detail.

In particular, most of the current EDF calculations utilize for the many-body wave function expansion either a mesh with a given size of the box and a distance between neighbor points, or a finite
number of harmonic oscillator single-particle states. Observables, like binding
energies, radii, etc., should in principle be independent of a particular choice for the working basis. Nonetheless, such an independence is only
obtained in calculations in a mesh if a sufficiently large and dense box is used. On the other hand, a large number of single-particle states have to be
included in the calculation with harmonic oscillator bases. This is rarely the case in practical applications due to limited computational resources. Hence, increasing the size of the working basis usually leads to an emergence of a convergence pattern for the calculated observables. In the case of calculations in a mesh, such convergence studies have been systematically performed recently with Skyrme functionals (see Ref.~\cite{PRC_92_064318_2015} and references therein). For harmonic oscillator bases, extrapolations
schemes to the limit of an infinite basis have been
used~\cite{1980PhRvC..21.1568D,2009PhRvL.102x2501G,2013PhRvC..88b4308G,2007EPJA...33..237H,2012PhRvC..86c1301F, 2013PhRvC..87d4326M,2014PhRvC..89d4301F,2015JPhG...42c4032F} as well as modifications of the basis in the so-called transformed harmonic oscillator method~\cite{PRC_58_2092_1998}.

One of the goals of this paper is to analyze the
convergence of energies computed with an underlying harmonic
oscillator single-particle basis using the variational HFB method. By doing this, we can directly test the global validity of the central ansatz for a widely implemented phenomenological extrapolation prescription in some of the previous large-scale HFB-based calculations~\cite{2007EPJA...33..237H,2009PhRvL.102x2501G,2010JPhG...37f4015B}. To our best knowledge, none of the earlier publications addressed the accuracy of this approach across entire isotopic chains. Having performed the convergence analysis, we turn our attention to a more recent extrapolation method, that was theoretically derived for calculations performed in harmonic oscillator basis. However, previous studies have evaluated the performance of this extrapolation strategy on a couple of simple systems for which ``exact'' many-body calculations are possible. In this paper, we introduce necessary tools and establish appropriate criteria for the systematic analysis of this extrapolation strategy applied to HFB calculations using Gogny EDF.

In the first part of the paper (Sec.~\ref{subsec:HFBmethod}) we briefly discuss the HFB method
and the general properties of the harmonic oscillator working basis (Sec.~\ref{subsec:HFBbasis}). In Sec.~\ref{sec:ConvAnalysis} we analyze the convergence patterns of HFB calculations with the variation of the numerical parameters of the basis. Then, we describe the most important
aspects of an extrapolation scheme introduced by Furnstahl, Hagen and
Papenbrock in
Ref.~\cite{2012PhRvC..86c1301F} and improved
subsequently in Refs.~\cite{2013PhRvC..87d4326M,2014PhRvC..89d4301F,2015JPhG...42c4032F}
(Sec.~\ref{L2-extrapolation_1}-\ref{L2-extrapolation_2}). In
Sec.~\ref{results_1}, this method is applied to the nucleus $^{16}$O
as a benchmark. This analysis is generalized to the nucleus $^{120}$Cd
and the cadmium isotopic chain in Sec.~\ref{results_2}, where we
identify the potential problems that could appear in the
extrapolation. Finally, the main results are summarized in
Sec.~\ref{summary}.

%
 \section{Theoretical framework}
 \label{sec:THEO}
%

  \subsection{Hartree-Fock-Bogolyubov (HFB) method}
  \label{subsec:HFBmethod}

The HFB method is based on the variational principle, where the variational many-body space is spanned by the product-type HFB wave functions~\cite{RingSchuck1980}
\begin{equation}
|\Phi\rangle = \prod\nolimits_{k}\hat{\beta}_{k} |0\rangle,
\label{eqn:HFBtype}
\end{equation}
with the property of being vacuum states with respect to the Bogolyubov quasiparticles, i.e.
\begin{equation}
\hat{\beta}_{k}|\Phi\rangle=0 \,\, \forall \,\, k.
\label{eqn:HFBvacuum}
\end{equation}
Bogolyubov quasiparticle creation and annihilation operators, $\hat{\beta}^{\dagger}_{k}$ and $\hat{\beta}_{k}$, are the most general linear transformation of arbitrary single-particle operators $\hat{c}^{\dagger}_{i}$ and $\hat{c}_{i}$~\cite{RingSchuck1980},
\begin{equation}
\hat{\beta}^{\dagger}_{k}=\sum\nolimits_{i}U_{ik}\hat{c}^{\dagger}_{i}+V_{ik}\hat{c}_{i},
\label{eqn:HFBtrans}
\end{equation}
where the matrices $U_{ik}$ and $V_{ik}$ are sought by the minimization of the total energy. Since the HFB states $|\Phi\rangle$ violate the particle-number symmetry, the minimization is performed with constraints on the desired expectation values of neutron and proton number operators $\hat{N}$ and $\hat{Z}$, so that $\langle\Phi|\hat{N}|\Phi\rangle=N$ and $\langle\Phi|\hat{Z}|\Phi\rangle=Z$. Hence, the HFB equations that define the ground-state $|\Phi_{0}\rangle$ are found by the condition:
\begin{equation}
\delta\left(E^{'}_{\mathrm{HFB}}\left[|\Phi\rangle\right]\right)_{|\Phi\rangle=|\Phi_{0}\rangle}=0
\label{eqn:HFBvar}
\end{equation}
with
\begin{equation}
E^{'}_{\mathrm{HFB}}\left[|\Phi\rangle\right]=\langle\Phi|\hat{H}-\lambda_{N}\hat{N}-\lambda_{Z}\hat{Z}|\Phi\rangle,
\label{eqn:HFBprime}
\end{equation}
where $\lambda_{N}$, $\lambda_{Z}$ are Lagrange multipliers to ensure the constraints above, while the $\hat{H}$ is the effective nuclear Hamiltonian. In the present study, the Gogny D1S interaction~\cite{1984NuPhA.428...23B} is used to define the energy density functional and the HFB equations are solved using the computer code developed at the Universidad Aut\'onoma de Madrid~\cite{axial_code} based on the gradient method. Here, all terms have been included in the Hartree-Fock (direct and exchange) and pairing fields except the pairing part from the spin-orbit term which is very small.

\subsection{Spherical harmonic oscillator single-particle
  basis}
\label{subsec:HFBbasis} 

\begin{figure}[t]
  \begin{center}
    \includegraphics[width=\columnwidth]{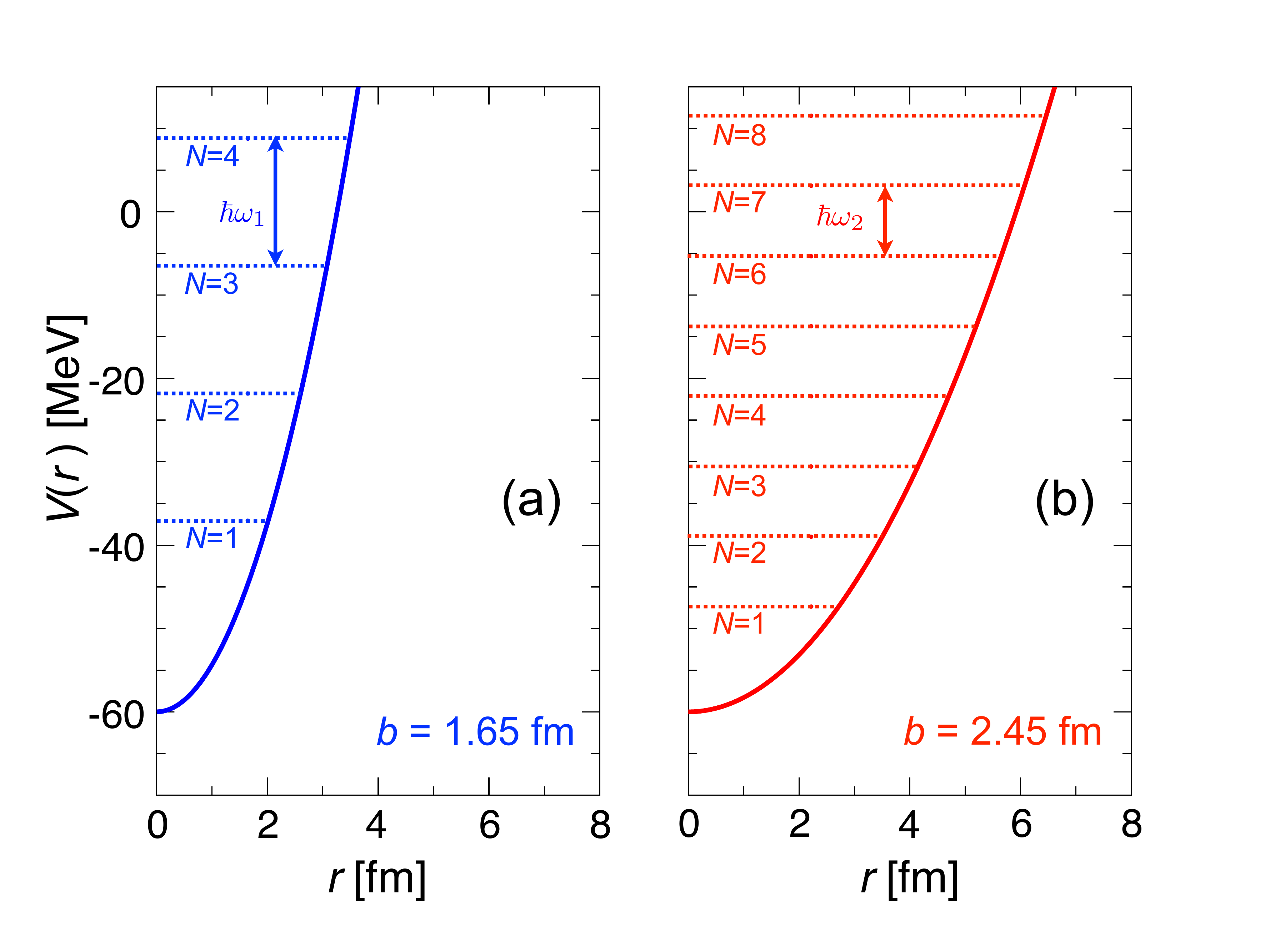}
    \caption{(color online) Spherical harmonic oscillator levels for two different values of the oscillator length (a) $b=1.65$ fm and (b) $b=2.45$ fm.}
    \label{Fig1}
  \end{center}
\end{figure}

A common choice of the single-particle working basis for the
quasiparticles expansion (Eq.~\ref{eqn:HFBtrans}) is a set of
spherical harmonic oscillator (SHO) functions. In this case there are
two numerical parameters that define the basis itself. The first one
is the total number of major oscillator shells included in the basis,
$N_{OS}$, which defines its dimension $d_{\text{tot}}$, i.e. the
number of single particle states, as
\begin{equation}
d_{\text{tot}}=\sum_{N=1}^{N_{OS}}D(N) = \frac{1}{3} N_{OS} (N_{OS}+1)(N_{OS}+2),
\label{eqn:SHOdim}
\end{equation}
where $D(N)=N(N+1)$ is the degeneracy of a single oscillator shell.
Here, $N=1,2,...$ is the major oscillator number $N=2n+l+1$, with $n=0,1,2,...$ and $l$
being the radial and angular momentum quantum numbers, respectively.

The second parameter of the basis is the intrinsic oscillator length $b$ of the SHO functions, which is connected to the oscillator energy $\hbar\omega$ as
\begin{equation}
b=\sqrt{\frac{\hbar}{m\omega}}.
\label{eqn:SHOb2hw}
\end{equation}

In Fig.~\ref{Fig1} we represent schematically the well-known spherical
harmonic oscillator potential, 
$V(r)=-V_{0}+(r/b)^2\hbar\omega/2$, for two different values of the
oscillator length $b=1.65$ fm and $2.45$ fm. The depth of the well is chosen to be the same for both schematic potentials, $V_{0}=60$ MeV. It is thus clear that for a
fixed number of $N_{OS}$, the maximum energy reached by a
single-particle state will be larger when the intrinsic oscillator length $b$ (and therefore the effective radius of the basis) is smaller. Nevertheless, both
bases are equivalent and should yield identical results for calculated observables when an infinite value of $N_{OS}$ is considered. However, due to basis truncations in practical calculations and an improper asymptotic behavior of the harmonic oscillator wave functions at long distances, such an independence from the numerical basis parameters ($N_{OS}$, $b$) is rarely reached.

%
    \section{Convergence analysis}
    \label{sec:ConvAnalysis} 
%

\begin{figure}[t]
  \begin{center}
    \includegraphics[width=1\columnwidth]{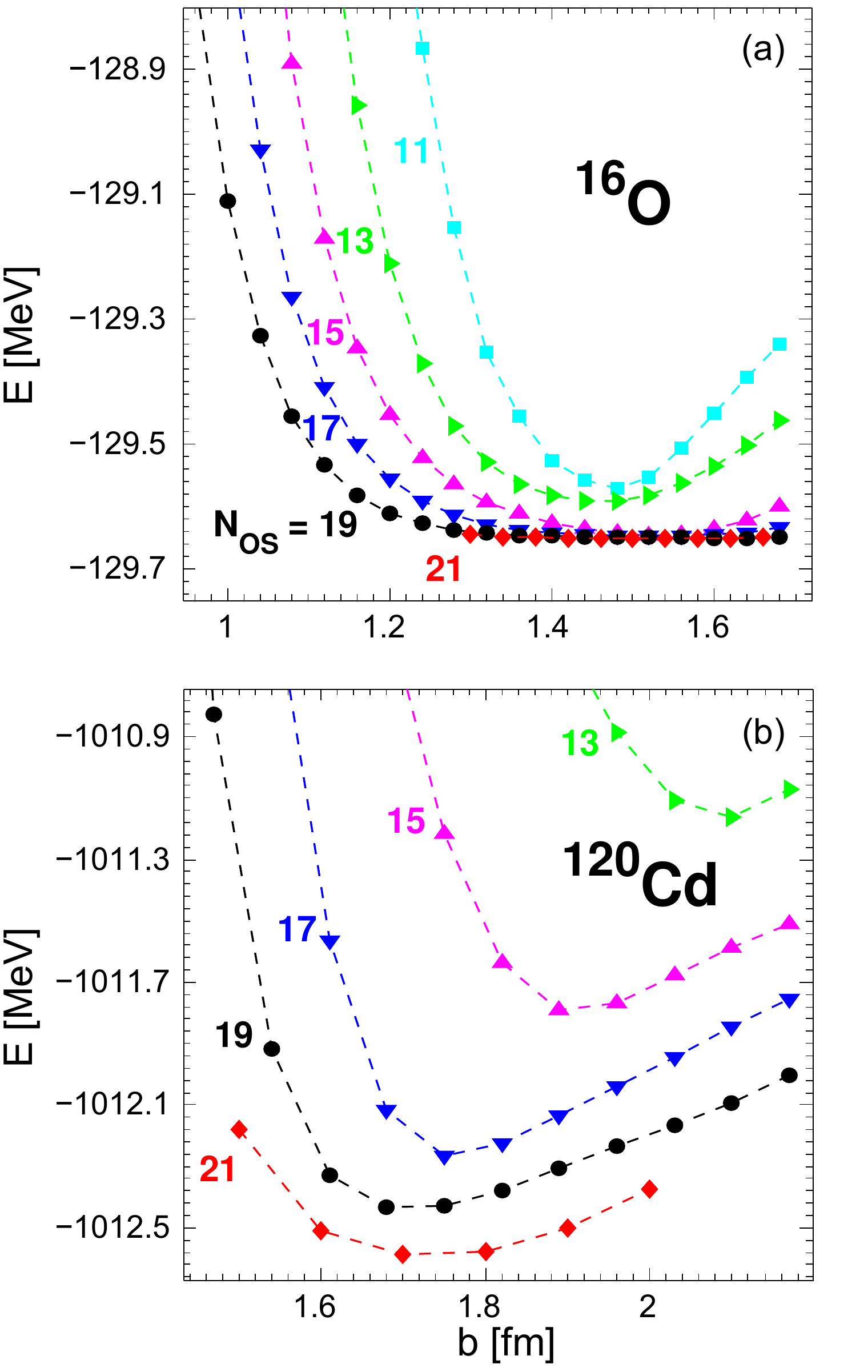}
    \caption{(color online) HFB energies calculated in different bases $N_{OS}=11$,...,$21$ (see labels at each curve) as a function of oscillator length $b$ for (a) $^{16}$O, and (b) $^{120}$Cd.}
    \label{Fig2}
  \end{center}
\end{figure}

Fig.~\ref{Fig2} shows the calculated ground-state (g.s.) HFB energies of $^{16}$O and $^{120}$Cd for bases $N_{OS}=11,...,21$ that are plotted against various oscillator length values $b$. One sees that going from $N_{OS}=11$ to $N_{OS}=13$, or from $N_{OS}=13$ to $N_{OS}=15$ yields noticeably deeper minima. Yet given a sufficiently large basis, g.s. energies of $^{16}$O nucleus are largely insensitive to the numerical parameters $N_{OS}$ and $b$, see Fig.~\ref{Fig2}(a). We can thus state that in this case the results are virtually converged to the true HFB energy, thereafter to be referred to as $E_\infty$. However, as was already mentioned, a complete convergence is rarely achieved in practice. For example, the calculated g.s. energies of the neutron-rich $^{120}$Cd in Fig.~\ref{Fig2}(b) are rather sensitive to the chosen intrinsic length of the basis $b$, even in larger bases with greater $N_{OS}$ values. Hence, further energy gain is anticipated from expanding the dimension of the working basis beyond our current maximum of 
$N_{OS}=21$.

\begin{figure}[t]
  \begin{center}
    \includegraphics[width=\columnwidth]{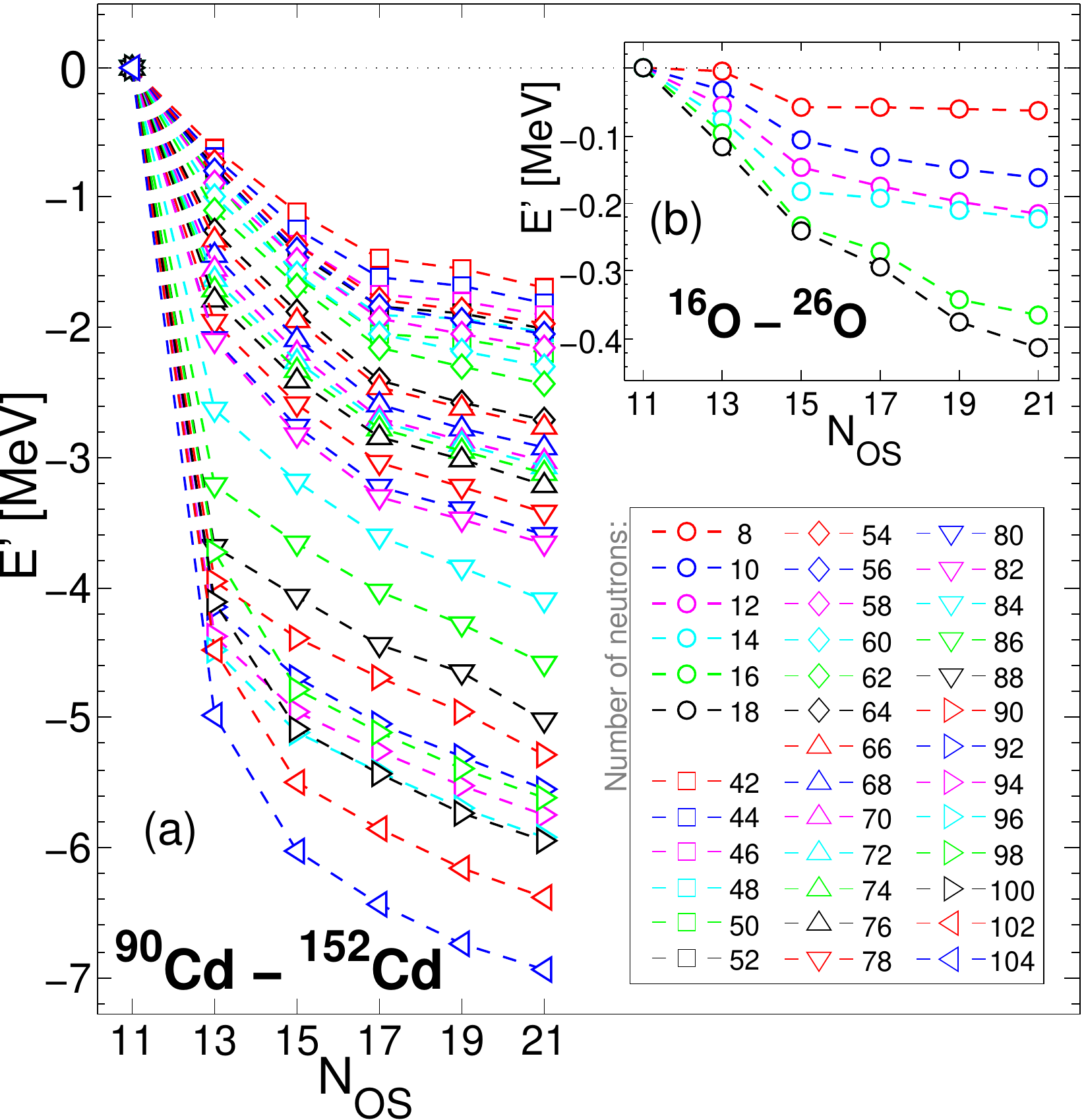}
    \caption{(color online) Convergence patters of HFB energies with enlargement of the basis dimension, defined as $E'(N_{OS}) = E_\mathrm{min}(N_{OS}) - E_\mathrm{min}(11)$ for (a) oxygen, and (b) cadmium nuclei.}
    \label{Fig3}
  \end{center}
\end{figure}

\begin{figure*}[t]
  \begin{center}
    \includegraphics[width=1\textwidth]{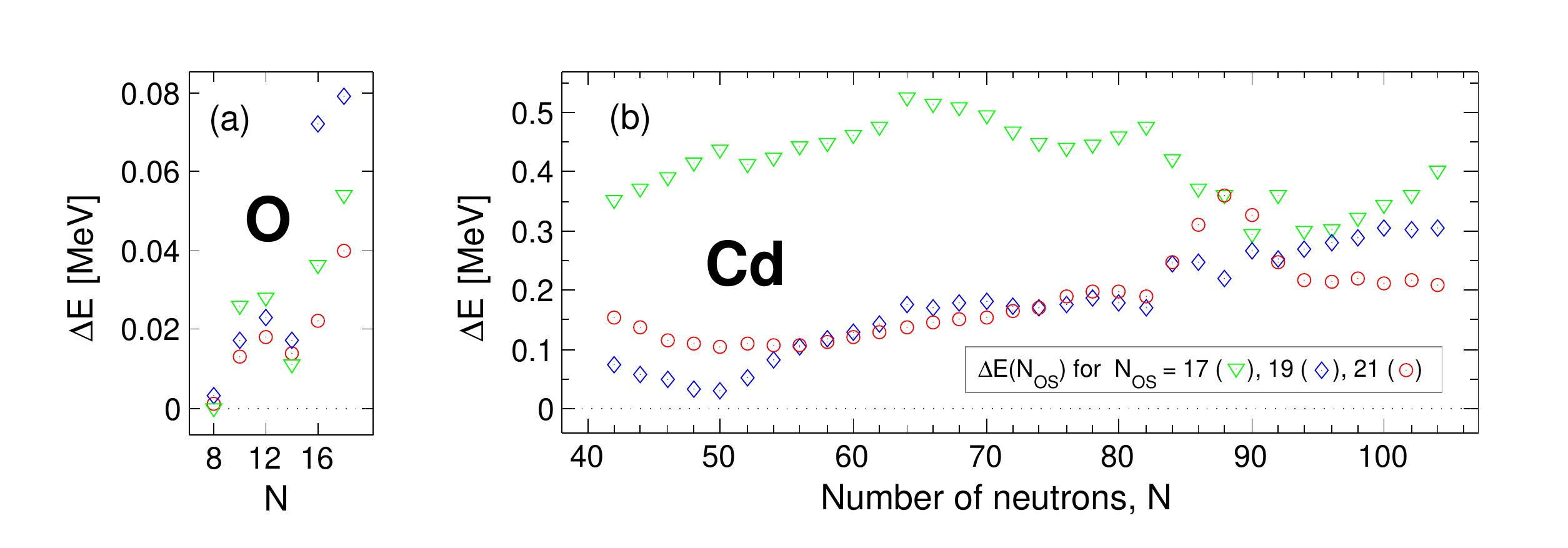}
    \caption{(color online) Obtained HFB energy gains by including two more major oscillator shells in the working basis, $\Delta E(N_{OS})=E_\mathrm{min}(N_{OS}-2)-E_\mathrm{min}(N_{OS})$, for (a) oxygen, and (b) cadmium isotopes.}
    \label{Fig3_1}
  \end{center}
\end{figure*}

We generalize these results to the study of g.s. energies in two isotopic chains, namely, oxygen and cadmium isotopes. In Fig.~\ref{Fig3} we show the energy gained by increasing the number of major harmonic oscillator shells with respect to the energy obtained with $N_{OS}=11$. Additionally, these values are calculated with the optimal choice of the oscillator length for each $N_{OS}$, 
\begin{equation}
E_\mathrm{min}(N_{OS})=\min\{E\mathrm(N_{OS}, b)\},
\label{eqn:Leff-reference}
\end{equation}
i.e., they correspond to the minima of the curves shown in Fig.~\ref{Fig2}. 
First of all, a flat behavior in the HFB energies with respect to $N_{OS}$ means a converged calculation. However, we see in Fig.~\ref{Fig3} that a strict convergence is reached only for the nucleus $^{16}$O. In the rest of the oxygen and cadmium nuclei we observe an increase in energy gain when we include more harmonic oscillator states in the working basis. Such an increase is larger for heavier isotopes. For example, performing calculations in a basis with $N_{OS}=21$ for $^{16}$O yields only $\sim$ 0.06 MeV of extra g.s. energy compared to a calculation with $N_{OS}=11$, and such gains gradually grow reaching $\sim$ 0.42 MeV for the drip line nucleus $^{28}$O. The situation with cadmium nuclei is similar, but the lack of convergence in the $N_{OS}=11$ basis is much more profound for these heavier systems. Hence, the calculation with $N_{OS}=11$ is underconverged by 1.70 MeV for $^{90}$Cd compared to the calculation with $N_{OS}=21$, and this value reaches 6.94 MeV for the nucleus $^{152}$Cd. In Fig.~\ref{Fig3} we 
also observe that the energy gain obtained by increasing the basis with two units of $N_{OS}$ is not always monotonic. To get more insight on this matter, we define such an energy gain as:
\begin{equation}
	\Delta E(N_{OS})=E_\mathrm{min}(N_{OS}-2)-E_\mathrm{min}(N_{OS}),
\label{eqn:DeltaE}
\end{equation}
%
%
Fig.~\ref{Fig3_1} shows $\Delta E(N_{OS})$ for $N_{OS}=17,19,21$ in
oxygen and cadmium isotopic chains. First, we see once again a fully
converged calculation for $^{16}$O with $\Delta E(N_{OS})$ effectively
being equal to zero. Second, we notice irregularities in
the convergence pattern for the majority of nuclei. This is
particularly well seen for the cadmium isotopes, where the convergence
pattern $\Delta E(21)<\Delta E(19)<\Delta E(17)$ does not generally hold. In
addition, we can also notice a clear disturbance of the slowly varying
$\Delta E(N_{OS})$ patterns in the isotopic region around the magic
$^{130}$Cd nucleus. Therefore, any extrapolation method that assumes
a continuous and smooth reduction of the energy gain obtained by
adding two major shells is not supported by the present
calculations~\cite{1980PhRvC..21.1568D,2007EPJA...33..237H,
  2009PhRvL.102x2501G,2010JPhG...37f4015B}.
  
%
\section{Extrapolations to an infinite basis}
\label{sec:EXTRAPOLATION}
%

The evident incomplete convergence in practical HFB calculations of
ground-state energies prompts us to search for a systematic and
reliable method to extrapolate the results obtained in a truncated
harmonic oscillator basis to the limit of an infinite basis. One of
the early attempts to quantify for the numerical error due to the
basis truncation is based on the assumption that the g.s. energy
follow a law $\Delta E(N_{OS}) \approx \Delta E(N_{OS}-2)/2$, which by summing the arithmetic series would imply an
estimate $E_\infty \approx E_\mathrm{min}(N_{OS})-\Delta E(N_{OS})$
~\cite{1980PhRvC..21.1568D,2007EPJA...33..237H,
  2009PhRvL.102x2501G, 2010JPhG...37f4015B}. According
to our previous discussion of Fig.~\ref{Fig3_1}, this ansatz is too
crude and not general enough to give a reliable estimation of
$E_\infty$. A number of more elaborate phenomenological extrapolation
schemes have also been used in nuclear structure
calculations~\cite{2009PhRvC..79a4308M,
  2008PhRvC..77b4301F, 2008NuPhA.801...21B,
  2007PhRvC..76d4305H}, but most of them include some
arbitrary aspects which prevent their use in global calculations.

The rest of the paper is devoted to the performance analysis of a more
general theoretically justified extrapolation scheme that was first introduced by Furnstahl, Hagen, and
Papenbrock in Ref.~\cite{2012PhRvC..86c1301F}, and
subsequently developed in Refs.~\cite{2013PhRvC..87d4326M,
  2014PhRvC..89d4301F,
  2015JPhG...42c4032F}. The underlying idea behind this approach is on a par
with assertions of quantum field theories, where the energy of a
particle enclosed in a finite volume is shifted by the imposed
boundary conditions. For example, it was shown in
Refs.~\cite{1986CMaPh.104..177L, 1986CMaPh.105..153L}
that the mass of a trapped particle exhibits an exponential
convergence to the infinite volume value at a certain theoretically
predicted rate. In our case, the
corresponding spatial confinement is present by virtue of the
localized nature of the SHO basis. The effective dimensions of the
enclosing volume are deduced from the spatial extensions of the
oscillator functions. By truncating our working basis, we effectively
impose a spherical hard-wall boundary condition in coordinate space
and an analogous intrinsic sharp boundary in momentum space. These
induced infrared (IR) and ultraviolet (UV) cutoffs of the basis, 
$\Lambda_\mathrm{IR}$ and $\Lambda_\mathrm{UV}$, are
modulated by the actual nucleus in consideration and the model space
parameters $N_{OS}$ and $b$, but are independent of the particular
potential used. With the cutoffs explicitly considered, it is possible
to derive the finite volume corrections to various nuclear structure
observables, such as g.s. energies and radii, hence effectively
extending the dimensions of the working basis to infinity. 

As a proof of theoretical concept, a row of successful tests for the suggested extrapolation were performed on a number of model potentials, as well as an example of the deuteron with a realistic chiral effective field theory ($\chi$EFT) potential~\cite{2012PhRvC..86c1301F}. Although derived at first only for systems that could be reduced to single-particle degrees of freedom, the extrapolations showed a good reliability and robustness even in many-body calculations. Hence, the extrapolation method was used in the nuclei $^{6}$He and $^{16}$O computed with a no-core shell model and a couple-cluster method respectively~\cite{2012PhRvC..86c1301F}. Since then, the extrapolation for the binding energy has also been applied to several other nuclei \cite{2013PhRvC..87e4312J, 2014PhRvC..90b4325R,2013PhRvC..87a1303S,2013PhRvC..87c4307H,2014PhRvC..89a1303S}, but without a particular analysis of its reliability.

Based on the previous insights, in
Ref.~\cite{2015JPhG...42c4032F}, Furnstahl et al. have
enhanced the theoretical basis of the derived IR
correction formula to extend its applicability
to many-body fermionic systems. The tests performed in three oxygen
isotopes, $^{16, 22, 24}$O, generally confirmed the anticipated
improvement of such IR extrapolations for atomic nuclei and brought us
closer to the question of error quantification of the extrapolation.

Despite the demonstrated success of the method for these individual
nuclei, the proposed scheme has not yet been put to a systematic test
with widely used EDFs, exploring its precision, accuracy and
reliability throughout the whole isotopic chains, particularly in the
neutron-rich extremes of heavier nuclear systems where the lack of
convergence is at largest. It is the purpose of this section to
systematically test the performance of the suggested energy correction
procedure within the HFB framework with the
Gogny D1S EDF. We start by introducing the relevant tools for the energy
extrapolation on an example of $^{16}$O. Later, we check the
performance of the method in the nucleus $^{120}$Cd. Finally, we
perform a systematic study of the
IR extrapolation scheme in the cadmium isotopic
chain from proton- to neutron-drip lines.

\subsection{Characteristic cutoffs of the basis}
\label{L2-extrapolation_1}

Following the arguments addressed in
Refs.~\cite{2012PhRvC..86e4002C,2012PhRvC..86c1301F}, there are two
momentum cutoffs imposed by the truncation of the model space for a
given finite single-particle basis of harmonic oscillator
functions. One of the cutoffs is associated with the highest
excitation energy of the chosen basis,
$E_{max}=\hbar\omega(N_{OS}+3/2)$ (see Fig.~\ref{Fig1}). In a
semiclassical approximation, the maximum momentum a particle in such a
basis can reach is $\Lambda_0 \equiv \sqrt{2mE_{max}}$ or in terms of
basis parameters
\begin{equation}
\Lambda_0 = \sqrt{2(N_{OS}+3/2)}\cdot \hbar / b.
\label{eqn:extrap:lamdba0}
\end{equation}
We take this as a leading-order estimate of the corresponding UV momentum cutoff of the basis, i.e., $\Lambda_{UV} \approx \Lambda_0$.

The second cutoff is induced in the opposite energy limit of the
finite SHO basis, which at low energies is shown to be effectively
equivalent to a spherical cavity with a sharp boundary radius
$L_\mathrm{IR}$~\cite{2014PhRvC..89d4301F}. Choosing
the classical turning point of a harmonic oscillator
$L_0 \equiv \sqrt{2 E_{max}/m\omega^2}$, or
\begin{equation}
L_0 = \sqrt{2(N_{OS}+3/2)} \cdot b,
\label{eqn:extrap:L0}
\end{equation}
as a first-order approximation for this radius suggests $L_\mathrm{IR} \approx L_0$. The associated IR cutoff is then defined as $\Lambda_\mathrm{IR} \equiv \pi / L_\mathrm{IR}$.

\begin{figure*}[t]
  \begin{center}
    \includegraphics[width=1\textwidth]{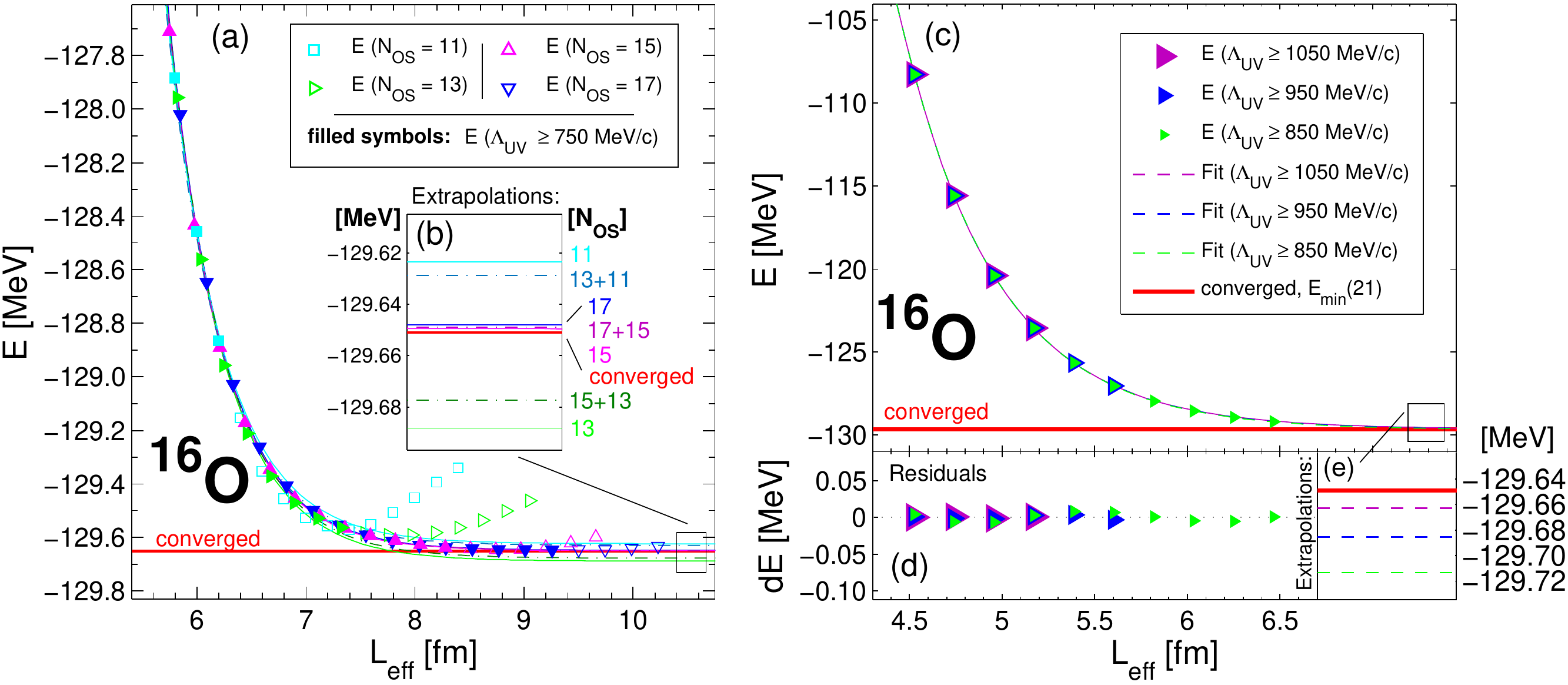}
    \caption{(color online) (a) calculated HFB energies for the nucleus $^{16}$O
      using SHO bases with $N_{OS}=11$,...,$17$ as a function of
      $L_\mathrm{eff}$. Filled symbols indicate g.s. energies having
      $\Lambda_{UV}>750$ MeV/c, red line shows the converged
      result, and other continuous and dot-dashed lines are the corresponding fits to
      Eq.~\ref{eqn:L2:1eFormula}; (b) inset shows $E_\infty$ values
      for different extrapolations; (c) HFB energies calculated in a
      basis having $N_{OS}=13$ and filtered out by conditions $\Lambda_\mathrm{thr}=850, 950,$ and
      $1050$ MeV/c. The corresponding fits are shown with dashed lines;
      (d) residuals defined as
      $dE=E(L_\mathrm{eff})-E_L(L_\mathrm{eff})$; and (e) the
      corresponding extrapolations. See figure labels.} 
    \label{Fig4}
  \end{center}
\end{figure*}

The complete convergence in a finite SHO basis can now be attained by
the fulfillment of both UV and IR conditions imposing constrains on
the choice of the basis. The first requirement is to select the basis
in such a way that the highest momentum scale $\lambda$ of the
employed interaction is smaller than the maximum momentum in the
working basis, i.e. $\lambda < \Lambda_{UV}$. This will ensure that
all the ultraviolet physics set by the interaction has been captured
in the working basis, which would provide a UV-converged results of
the calculation. The second condition requires that the effective
spatial radial extent $L_\mathrm{IR}$ of the chosen basis is large
enough to encompass the many-body wave function, i.e. $r < L_{IR}$. It
is this second condition that can usually never be fully achieved in
practice for neutron-rich nuclei due to the different asymptotic
behavior of the nuclear wave function (exponential falloff) and the
SHO basis (Gaussian falloff) in coordinate space. Thus, in order to
obtain the greatest degree of convergence in a truncated model space,
one usually performs calculations in the largest accessible $N_{OS}$,
and seeks for an optimal compromise between the IR and UV conditions
by finding the binding energy minimum through variation of the intrinsic
oscillator length $b$ (see Fig.~\ref{Fig2}). However, selecting
calculations performed only with sufficiently small oscillator
lengths, one can strive to ensure the UV condition and thereby
effectively isolate the systematic error coming from the lack of IR
convergence. 

While this is easily achievable in many-body calculations with
interactions where the cutoff is set using an UV-regulator, the
situation is different in the current EDF approaches. Since the Gogny
interaction has contact terms in the spin-orbit and density-dependent
part of the functional, it does not have an intrinsic momentum
cutoff. A. Rios and R. Sellahewa~\cite{Rios_Gogny} have recently shown
that the D1S parametrization, once decomposed in partial waves,
contains significant matrix elements connecting high and low momenta
in some channels of the interaction. Nonetheless, it still remains to
be investigated, whether these two-body matrix elements have
noticeable impact on the whole HFB calculation for a particular
nucleus under consideration. However, in many cases we are able,
\textit{a posteriori}, to determine the parameters of the working basis
in order to effectively ensure the UV criterium
$\lambda < \Lambda_{UV}$, whereupon the IR extrapolation scheme could
be applied to account for the IR corrections.

\subsection{The first-order IR--extrapolation}
\label{L2-extrapolation_2}

One of the actual challenges in accounting for the boundary effects
enacted by the IR-cutoff was the determination of the effective
impenetrable extend of the chosen set of SHO basis functions in a most
accurate and universal way. The choice of the maximum displacement,
$L_0$, can qualitatively explain the concept of extrapolation, but it
is only a leading-order estimate for the extent of the oscillator wave
function. As it was recently shown, the correct box size of the SHO
basis for many-body system is deduced by matching the smallest
eigenvalue of total squared momentum operator for a particular nucleus
in a given SHO basis to the analogous smallest value in the spherical
cavity~\cite{2015JPhG...42c4032F}. The
resulting effective radius $L_\mathrm{eff}$ is then 
\begin{equation}
	L_\mathrm{eff} = \left( \frac{ \sum\nolimits_{nl} \nu_{nl}a^{2}_{ln} } { \sum\nolimits_{nl} 		\nu_{nl}\kappa^{2}_{ln} }\right)^{1/2},
\label{eqn:extrap:Leff}
\end{equation}
where the $\kappa^2_{ln}$ are the eigenvalues of the momentum squared operator diagonalized in the SHO basis, $\nu_{nl}$ are the occupation numbers of nucleons giving the lowest kinetic energy in SHO basis; and $a_{ln}$ are the $(n+1)^\mathrm{th}$ zero of the spherical Bessel function $j_l$.

With the effective hard-wall boundary of the SHO basis properly identified, one can now recast the initial problem of having the given many-body system enclosed by a harmonic oscillator soft-cavity into the one with a sharp infinite potential with an effective radius $L_\mathrm{eff}$. Such problems of confined quantum systems have been already studied (e.g. \cite{1987JMP....28.1813F} and citations therein) with various techniques available for the energy corrections. One can proceed by making a linear energy approximation of the many-body wave function and impose a vanishing Dirichlet boundary conditions at $L_\mathrm{eff}$. Whereas the details of the derivation can be found in Ref.~\cite{2015JPhG...42c4032F}, the resulting analytical expression of the first-order IR-correction is then of the form 
\begin{equation}
  E_L(L_\mathrm{IR}) = A_\infty \exp{(-2k_\infty L_\mathrm{IR})} + E_\infty,
  \label{eqn:L2:1eFormula}
\end{equation}
where for the atomic nucleus the proper radius is $L_\mathrm{IR} = L_\mathrm{eff}$ that depends both on the basis and the particular isotope, while $A_\infty$, $k_\infty$, and $E_\infty$ are taken as fit parameters. This derived exponential pattern of the IR correction was shown to be independent of the particular potential and validated in the examples mentioned above ~\cite{2013PhRvC..87d4326M,2014PhRvC..89d4301F,2015JPhG...42c4032F}.

  \subsection{Playground test with $^{16}$O}\label{results_1}

\begin{figure}[t]
  \begin{center}
    \includegraphics[width=\columnwidth]{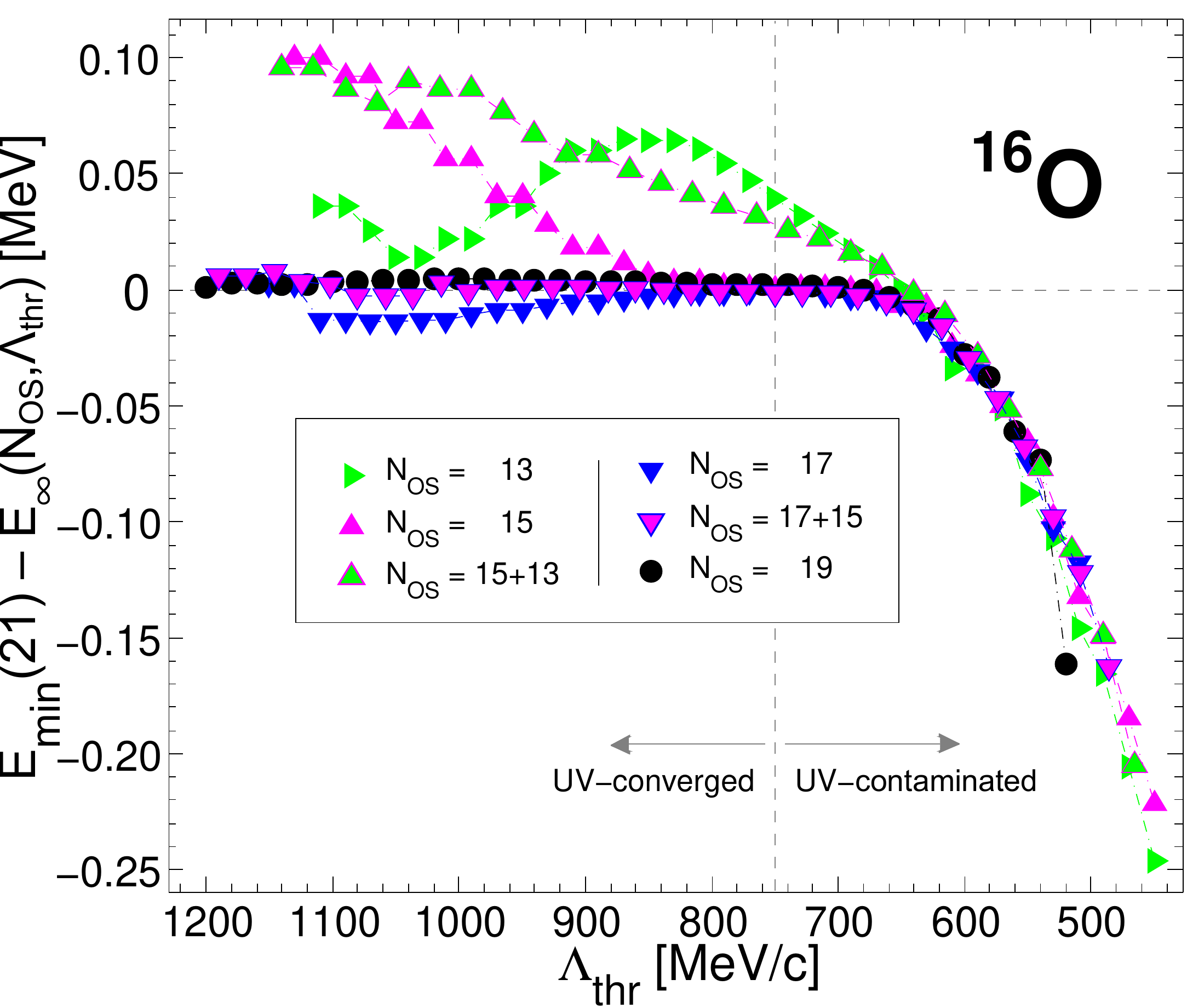}
    \caption{(color online) Difference of the calculated $E_\mathrm{min}(21)$ and the energy obtained with IR extrapolations from different combinations of basis dimensions $N_{OS}$ and $\Lambda_{thr}$ values.}
    \label{Fig5}
  \end{center}
\end{figure}

We now illustrate the suggested extrapolation concept and introduce
the relevant benchmarking tools for this method using the nucleus
$^{16}$O as an example. The commencing test with this nucleus is
prompted by the well-converged HFB results starting already with a
basis of $N_{OS}=15$, as is evident from Figs.~\ref{Fig2}
and~\ref{Fig3}. In Fig.~\ref{Fig4}(a)~and~(c) we show the HFB energy
as a function of the effective radial extent $L_\mathrm{eff}$. In
order to apply the IR corrections, we start by selecting only those
calculations for which the UV cutoff of the basis is sufficiently
large, so that the results are considered as UV converged. This is
done by taking into account only those g.s. energies that are computed
in a basis with the UV cutoff above a certain threshold value, that is
$\Lambda_{UV}>\Lambda_{thr}$. For this illustration we take
$\Lambda_{thr} = 750$ MeV/c, and justify this choice later. The
selected HFB energies are represented by the filled symbols in
Fig.~\ref{Fig4}(a). We find that all of them almost perfectly fall on
an exponential curve, consistent with the theoretical predictions for
UV-converged results. The observed rise of the g.s. energies at larger
values of $L_\mathrm{eff}$ (Fig.~\ref{Fig4}(a), hollow symbols) is due
to an insufficient UV convergence. Those calculations are excluded
from the fit to the form of Eq.~\ref{eqn:L2:1eFormula}.  The solid
lines in Fig. \ref{Fig4}(a) represent the separate fits to the HFB
energies calculated in different combinations of the basis dimensions,
$N_{OS}$. The inset, Fig. \ref{Fig4}(b), shows the corresponding
extrapolated values $E_\infty$ together with a reference value of a
virtually converged calculation $E_\mathrm{min}(21)$. We see in this
figure that differences around 60 keV are obtained. Although we do not
directly attribute such a spread in the extrapolated values to the
uncertainty of the method, it is nevertheless representative of the
precision and accuracy level of the extrapolation scheme.

Of course, an accurate, precise and reliable extrapolation should also
be largely insensitive to the choice of the threshold momentum
$\Lambda_{thr}$, as long as the UV-convergence is ensured. We verify
this criterion for $^{16}$O by fitting to different sets of HFB
calculations, defined by the choice of a threshold value
$\Lambda_{thr}=850,950,1050$ MeV/c. Moreover, in order to imitate a
typical situation (common to heavy and neutron-rich nuclear systems)
of having access only to an insufficiently large working basis for
complete convergence, we limit ourselves to a SHO basis with
$N_{OS}=13$. In this case the calculations for $^{16}$O are not fully
converged. The illustration of this benchmark can be seen in
Fig.~\ref{Fig4}(c), where fits for different threshold values are
provided by the colored lines. All HFB points are found to be on an
exponential curve and the quality of the exponential convergence
pattern is particularly well seen in Fig.~\ref{Fig4}(d). The
corresponding $E_\infty$ estimates of the fits, shown in
Fig.~\ref{Fig4}(e), yield a narrow spread of their values, falling
very close to the converged energy value $E_\mathrm{min}(21)$, thereby
indicating a good stability, accuracy and precision of the method in
this specific example.

We now perform a systematic analysis to estimate the dependence of the
extrapolated values on the choice of $\Lambda_{thr}$. It is expected
that below a certain value of $\Lambda_{thr}$, the computed HFB
energies values could be affected by a lack of UV convergence (or
'UV-contamination'). The knowledge of a lower limit of $\Lambda_{thr}$
will allow us to include as many computed HFB data points into our
extrapolation as possible. To this end, we perform a series of
extrapolations obtained in bases with various sets of $N_{OS}$ and $b$
parameters, and vary the threshold momentum across a wide range of
$450 \leq \Lambda_{thr} \leq 1250$ MeV/c. In Fig. \ref{Fig5} we plot
the difference between $E_\mathrm{min}(21)$ and the extrapolated
values. Hence, positive (negative) values give extrapolated
g.s. energies below (above) $E_\mathrm{min}(21)$ that is considered as
the converged g.s. energy.  We observe first that lowering
$\Lambda_{thr}$ below a certain limit, namely, 620 MeV/c, we start to
incorporate into our extrapolation an increasing amount of points
which are not sufficiently UV-converged. Therefore, the inclusion of
these points deteriorates the quality of the fit and should be
eliminated from the IR extrapolating data set. Resting upon the
results of these calculations, we estimate the threshold for a
significant UV-contamination lies around $\Lambda_{thr} \approx 750$
MeV/c in $^{16}$O. We also observe a slight dependence of the
extrapolated g.s. energies on $\Lambda_{thr}$ of about 0.1 MeV if HFB
results with $N_{OS}\leq15$ are considered, even in the regions well
above the estimated onset of the UV-contamination. Consequently, the
extrapolated results are not completely free of $\Lambda_{thr}$
dependencies unless a sufficiently large value of $N_{OS}$ is chosen.


To conclude this section, we summarize the necessary criteria for the IR extrapolation to be robust and reliable. Assuming that the set of parameters $(N_{OS},b)$ defining the basis of HFB calculations ensures the UV-convergence of the g.s. energies, the following properties must hold for the $E_\infty$ estimates:
\begin{enumerate}[i.]
\item independence of the chosen threshold value $\Lambda_{thr}$ that define the set of HFB energies used in the fits according to criteria $\Lambda_{UV}>\Lambda_{thr}$;
\item insensitivity to the basis dimensionality used to compute the HFB energies chosen in the fit dataset. That is, the $E_\infty$ values should be independent of whether we pick a calculation performed with $N_{OS}=17$, with $N_{OS}=19$, or even if we combine the two sets;
\item finally, given that the fully converged value of the HFB g.s. energy is generally unknown, extrapolations should be able to at least reproduce the best converged HFB calculation available, i.e. $E_\mathrm{min}(21)$ in this work, or yield $E_\infty$ estimates that are below that value.
\end{enumerate}

  \subsection{Cadmium isotopic chain}\label{results_2}

\begin{figure}[t]
  \begin{center}
    \includegraphics[width=\columnwidth]{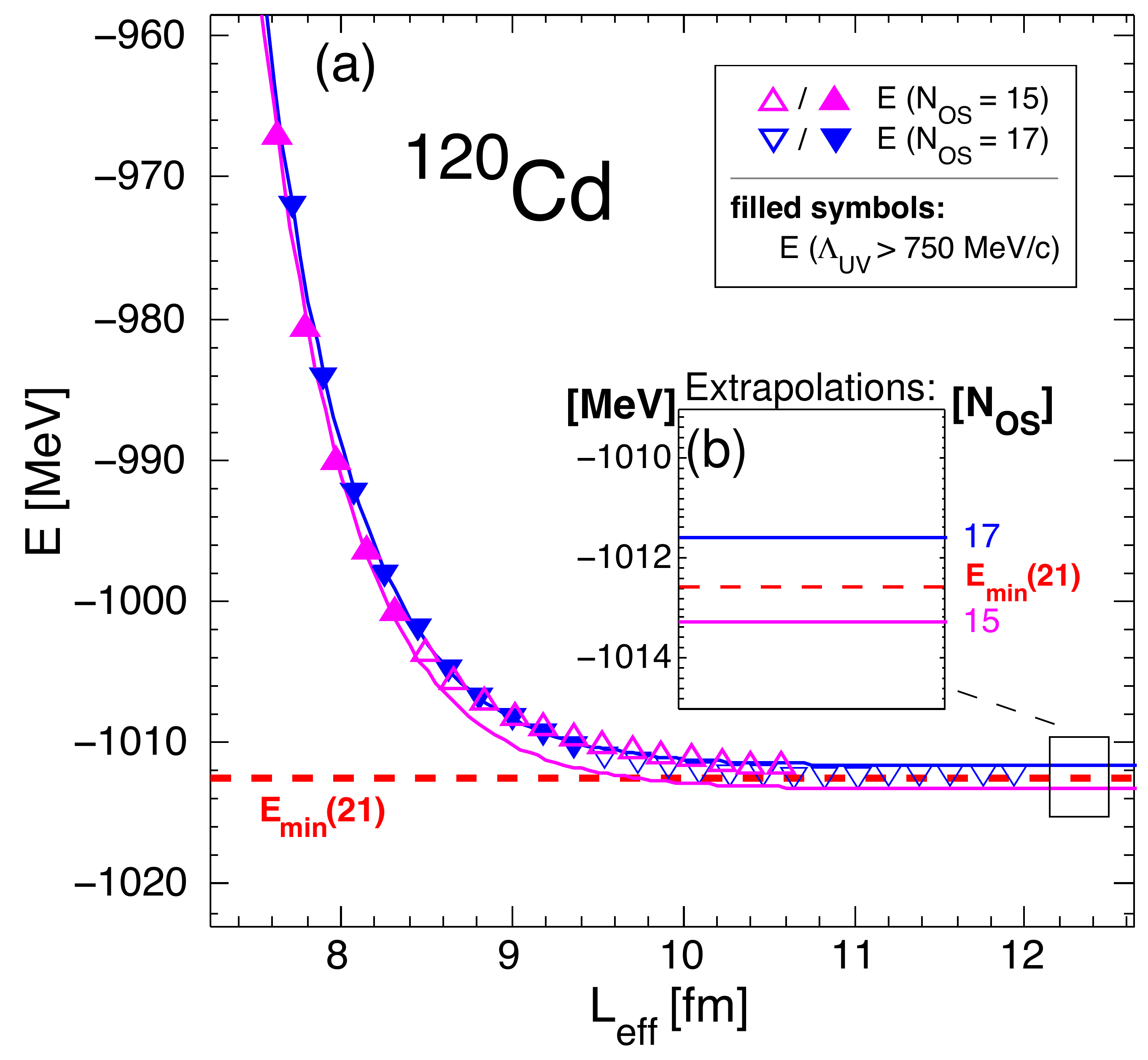}
    \caption{(color online) Similar to Fig.~\ref{Fig4} (a,b), but for $^{120}$Cd nucleus for $N_{OS}=15$ and $17$ bases. The $E_\mathrm{min}$ values is indicated by the dashed red line.}
    \label{Fig6}
  \end{center}
\end{figure}
\begin{figure}[b]
  \begin{center}
    \includegraphics[width=\columnwidth]{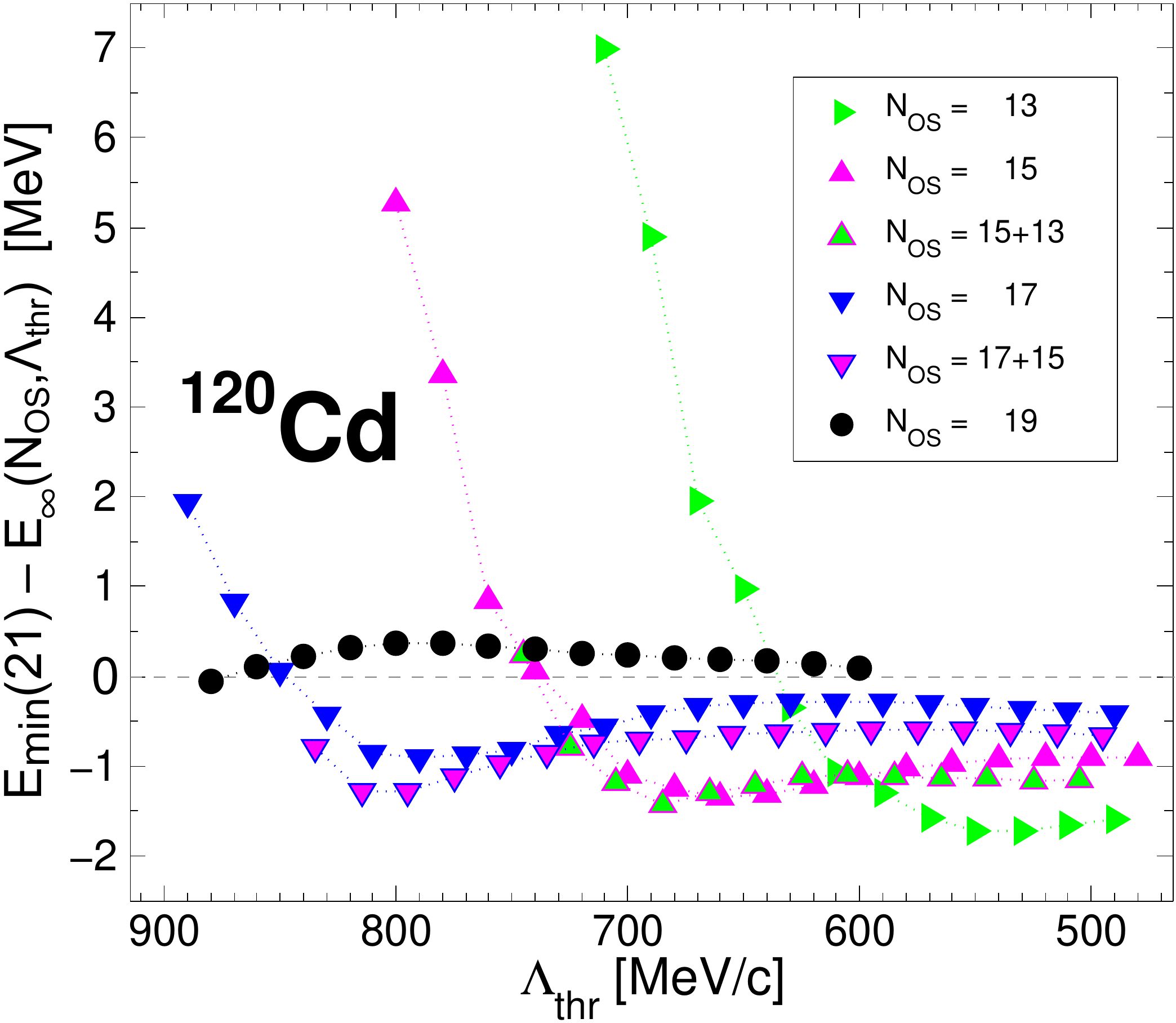}
    \caption{(color online) Similar to Fig.~\ref{Fig5} but for $^{120}$Cd.}
    \label{Fig7}
  \end{center}
\end{figure}
In the previous section we have studied the nucleus $^{16}$O to benchmark the IR extrapolation scheme and establish the main properties that the extrapolated energy should fulfill. We now apply the same method to extract the $E_\infty$ estimates first in the nucleus $^{120}$Cd, and then for the whole cadmium isotopic chain. As we showed in Figs.~\ref{Fig2}(b) and~\ref{Fig3}(a), none of these nuclei is fully converged. The HFB energy as a function of the effective spatial radial extent \textit{$L_\mathrm{eff}$} and the corresponding fits to Eq.~\ref{eqn:L2:1eFormula} for the nucleus $^{120}$Cd are plotted in Fig.~\ref{Fig6}. Following the prescription found in the previous section, we impose a cutoff of $\Lambda_{UV}>750$ MeV/c to select SHO bases with sufficiently high momentum cutoff. We observe in Fig.~\ref{Fig6}(a) a qualitative exponential decay with respect to \textit{$L_\mathrm{eff}$}. However, the extrapolated 
values show a larger spread in absolute energy than in the case of $^{16}$O, as can be seen in Fig.~\ref{Fig6}(b). For example, $E_\infty$ estimate is about $1.7$ MeV lower with the extrapolation from the $N_{OS}=15$ basis than from the one with $N_{OS}=17$. In addition, the minimal g.s. HFB energy attained in $N_{OS}=21$ basis, i.e. $E_\mathrm{min}(21)$ value, lies in between the two extrapolated energies mentioned above. Similarly to Fig.~\ref{Fig5}, we plot the dependence of the extrapolated energies on the $\Lambda_{thr}$ value in Fig.~\ref{Fig7} for the nucleus $^{120}$Cd. In this case the situation is far from fulfilling the requirements for a robust and reliable extrapolation given in the previous section. We found rather unstable results for extrapolations from $N_{OS}=13$, $15$, and $17$ bases for large values of $\Lambda_{thr}$ that are precisely the ones that should be better UV-converged. In those cases, the spread in the extrapolated energies produced by the particular choice of $\Lambda_{thr}$ can be as large as 7 MeV. In addition, when the fits are performed 
for the $N_{OS}=13$, $15$, and $17$ results separately, as well as combinations thereof, the extrapolated energies are systematically less bound than the best value reached with $N_{OS}=21$ basis in the range of $\Lambda_{thr}$ where a flatter behavior is found. Since the extrapolation method is intended to estimate the remaining energy missed by the truncation of the working basis, these results are not acceptable. However, the only exception are the extrapolations from HFB energies obtained with $N_{OS}=19$ basis, which seem to be most reliable.  

\begin{figure}[b]
  \begin{center}
    \includegraphics[width=\columnwidth]{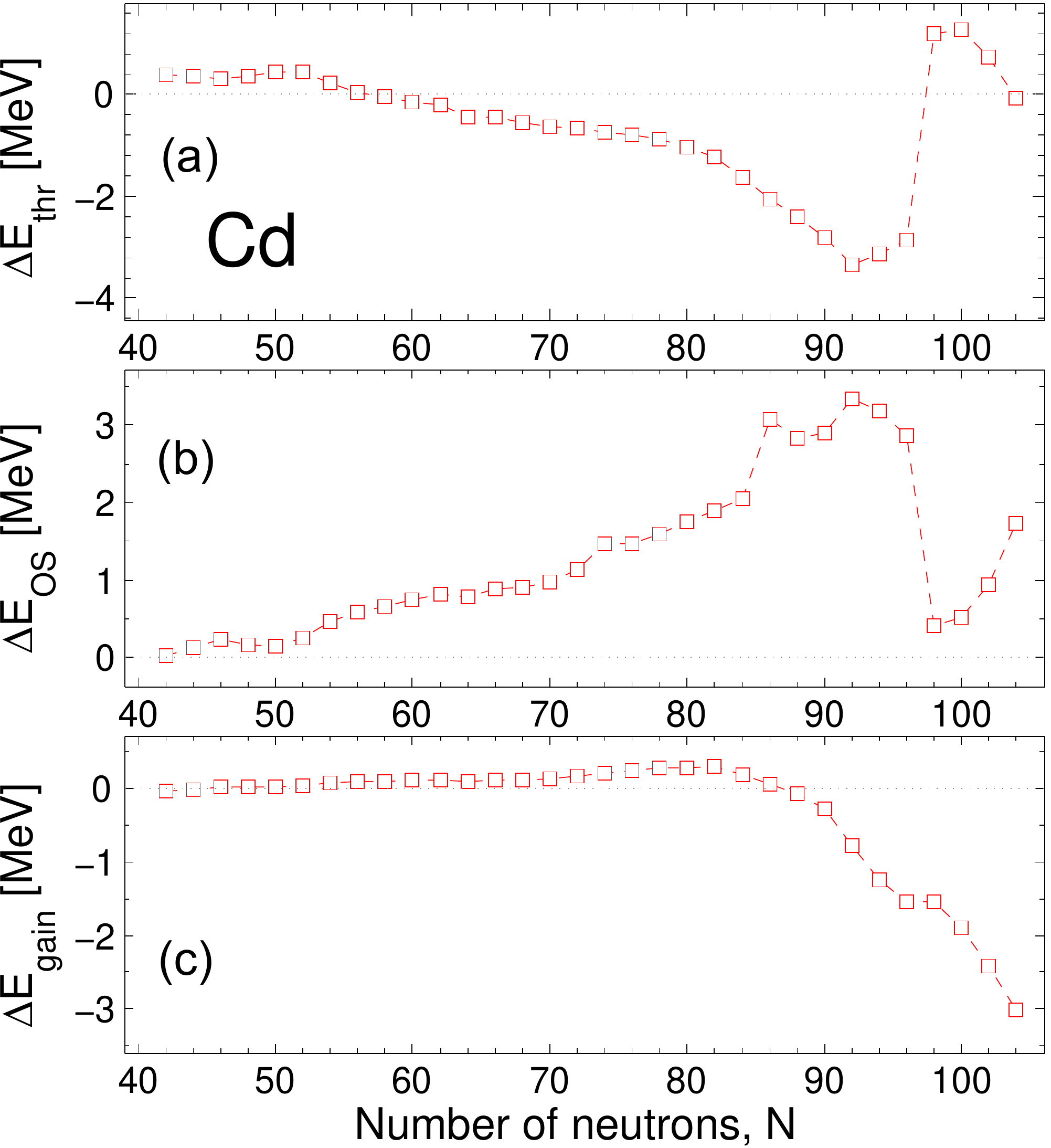}
    \caption{(color online) $\Delta E_{thr}$, $\Delta E_{OS}$, and $\Delta E_{gain}$ for $^{90-152}$Cd isotopes as a function of the neutron number.}
    \label{Fig8}
  \end{center}
\end{figure}

So far we have discussed the performance of the IR extrapolation method for individual nuclei. For the sake of completeness, we analyze the reliability and stability of the method in the whole cadmium isotopic chain. According to the points raised in the previous section to define the quality of the extrapolated energies, let us define the following quantities for each nucleus in the chain:
\begin{enumerate}[i.]
\item $\Delta E_{thr}\equiv E_{\infty}(\Lambda_{thr}=750\,\mathrm{MeV/c})-E_{\infty}(\Lambda_{thr}=900\,\mathrm{MeV/c})$ with $N_{OS}=19$ fixed to check the dependence on the chosen threshold value $\Lambda_{thr}$. Hence, $\Delta E_{thr}\approx0$ would mean a good performance;
\item $\Delta E_{OS}\equiv E_{\infty}(N_{OS}=17)-E_{\infty}(N_{OS}=19)$ with $\Lambda_{thr}=750\,\mathrm{MeV/c}$ fixed to check the dependence on $N_{OS}$. As in the previous point, $\Delta E_{OS}\approx0$ would mean a good performance;
\item $\Delta E_{\mathrm{gain}}\equiv E_{\mathrm{min}}(21)-E_{\infty}(N_{OS}=19,\Lambda_{thr}=750\,\mathrm{MeV/c})$ to check the quality of the extrapolation with respect to the lowest HFB energy computed in this work. Thus, $\Delta E_{\mathrm{gain}}$ should be equal to zero for converged cases and slightly positive for those HFB calculations which are not converged.
\end{enumerate}
In Fig.~\ref{Fig8} we show these three quantities as a function of the number of neutrons in the nuclei $^{90-152}$Cd. We observe first that the three conditions given above are not completely fulfilled simultaneously throughout the whole cadmium isotopic chain. Nevertheless, in many nuclei the dependencies on $\Lambda_{thr}$ and $N_{OS}$ are rather mild with $|\Delta E_{OS}|\approx|\Delta E_{thr}|\leq2$ MeV in the range of $N=42-84$. In this region, the differences of the extrapolated energies with respect to the best values obtained with $N_{OS}=21$ basis are close to zero or slightly above, providing a physically sound extrapolation. However, the situation is different in the neutron rich region above $N\geq86$, where the extrapolations are remarkably dependent on the choice of both $\Lambda_{thr}$ and $N_{OS}$, as well as lie above the best HFB energies directly computed, i.e., $\Delta E_{\mathrm{gain}}<0$. Therefore, whereas in the first region some systematic error bars could be extracted from the extrapolation, that is not the case in the neutron-rich region.

\begin{figure}[t]
  \begin{center}
    \includegraphics[width=\columnwidth]{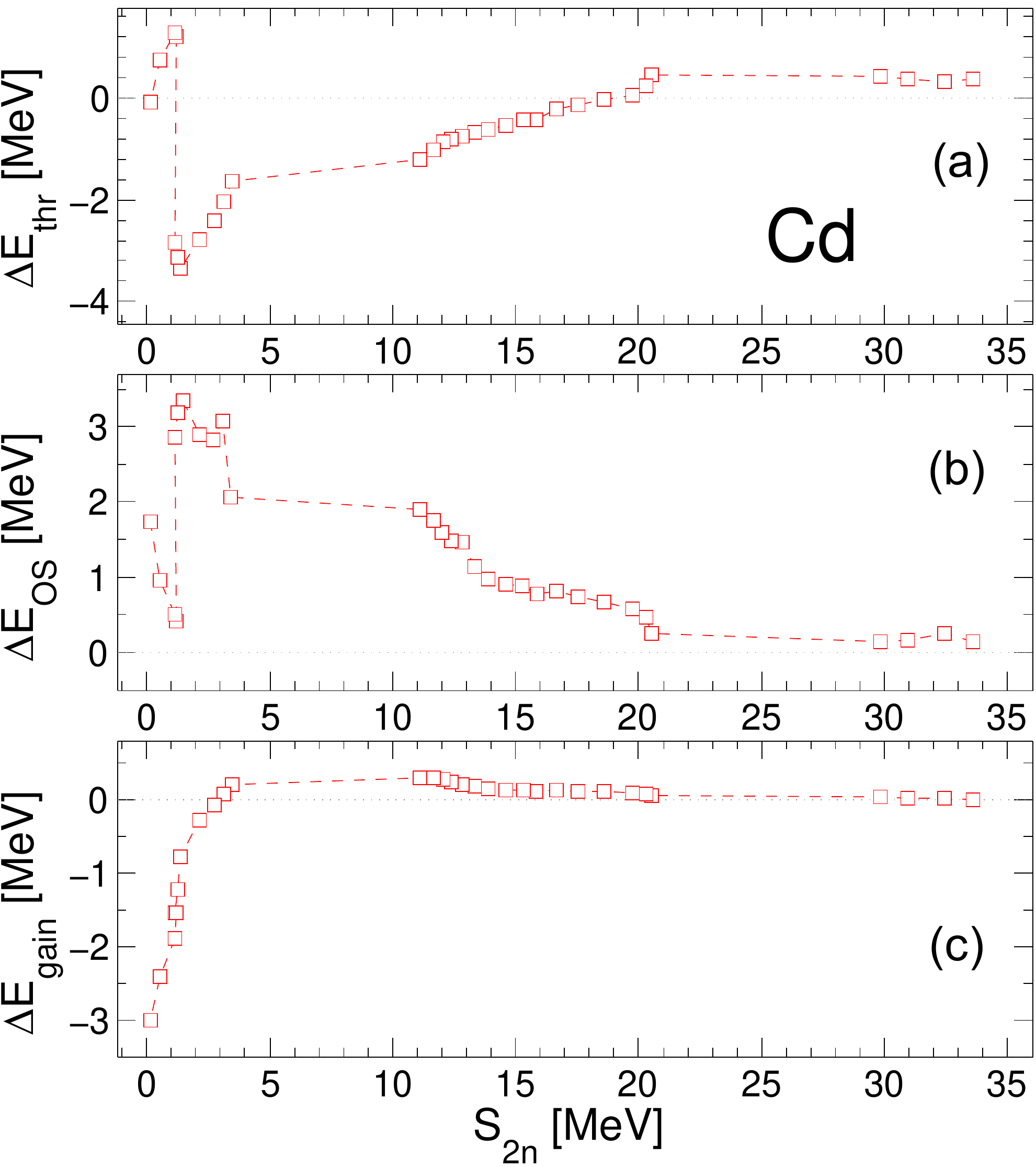}
    \caption{(color online) Similar to Fig.~\ref{Fig8} but as a function of the $S_{2n}$, which were obtained directly from HFB calculations in $N_{OS}=21$ basis \textit{without} extrapolations.}
    \label{Fig9}
  \end{center}
\end{figure}

We now represent in Fig.~\ref{Fig9} the same quantities but as a
function of the two-neutron separation energy $S_{2n}(N)\equiv E(Z,N-2)-E(Z,N)$, where the shell gaps, corresponding to $N=50$ and $82$ magic numbers, are well seen. From the astrophysical point of view, the most interesting aspect is that
the ill-behavior of the extrapolation method is significantly larger for isotopes beyond $N=82$
when the two-neutron separation energy is less than 5 MeV
approximately, which is precisely the relevant range in r-process calculations. 

\begin{figure}[t]
  \begin{center}
    \includegraphics[width=1\columnwidth]{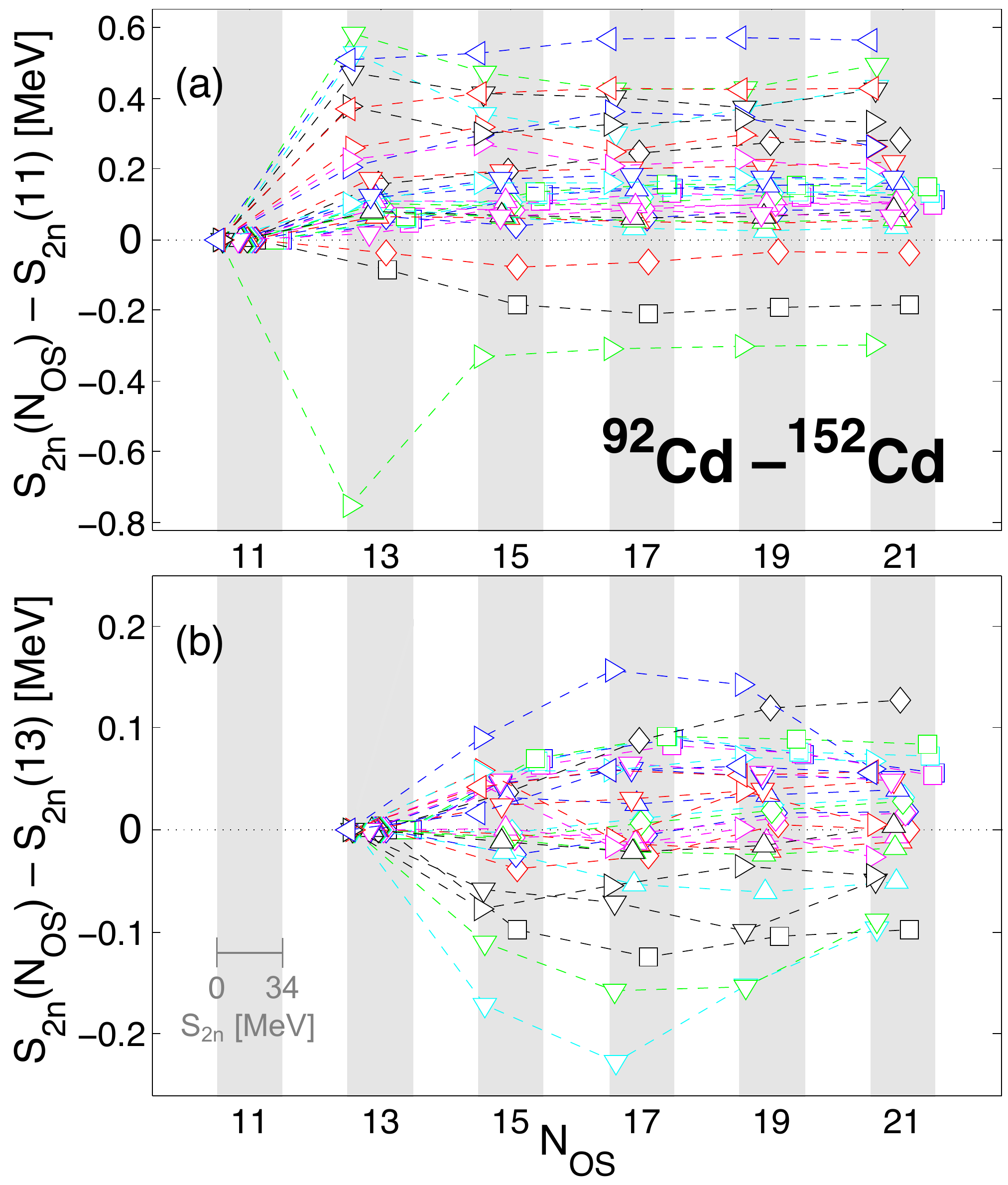}
    \caption{(color online) $S_{2n}$ energies of cadmium isotopes calculated in different basis dimensions (shaded bins). Color code and symbols are same as in Fig.~\ref{Fig3}.}
    \label{Fig11}
  \end{center}
\end{figure}

A similar result is obtained in the magnesium isotopic chain (not
shown) where such an erratic behavior of the extrapolated energy is
also found in the neutron rich part of the chain ($S_{2n}\leq5$
MeV). Therefore, the present extrapolation scheme is not able to
provide reliable estimations of $E_{\infty}$ values in those loosely bound regions, where the lack of convergence is also the largest. By the same arguments, the considered extrapolation scheme cannot be used to extract the two-neutron separation energies from the $E_{\infty}$ values. Despite this, being energy differences of the neighboring nuclei, the particle separation energies are expected to be better converged. Indeed, this is the case as can be seen in Fig.~\ref{Fig11}, where calculated $S_{2n}$ values are shown without any extrapolations. The obtained energies are distributed among the shaded column bins according to the basis dimensions of the calculations. Moreover, in order to see the convergence patterns more clear, the separation energies for each isotope are shifted down by a constant that equals to the $S_{2n}$ value obtained in $N_{OS}$=11 (Fig.~\ref{Fig11}(a)), or in $N_{OS}$=13 (Fig.~\ref{Fig11}(b)) basis. Furthermore, for better readability, the two-neutron separation energies are also displaced within each column bin so that the lower absolute $S_{2n}$ values are shifted closer to the left edge of each shaded region, while the higher ones are closer to the right edge (by analogy to Fig.~\ref{Fig9}). We see that for isotopes having $S_{2n}>5$ MeV, the enlargement of the basis beyond $N_{OS}=11$ affects their values up to 0.3 MeV at most, Fig.~\ref{Fig11}(a). For nuclei which have $S_{2n}<5$ MeV the spread around zero reference value is about double as high, reaching as much as 0.6 MeV for the dripline isotope $^{152}$Cd predicted by Gogny D1S EDF. By the same token, Fig.~\ref{Fig11}(b) shows the convergence patterns zeroed out for a somewhat larger $N_{OS}=13$ basis. Here we see that almost all of the $S_{2n}$ values in $N_{OS}=21$ basis fall within 0.1 MeV spread. However, we observe somewhat larger spread for nuclei having $S_{2n}<5$ MeV in $N_{OS}=15, 17$, and $19$ bases. All in all, despite the fact that two-neutron separation energies do not exhibit any noticeable convergence pattern when the basis size is increased, these quantities reach a much better degree on convergence already in relatively small bases.
\begin{figure*}[ht]
  \begin{center}
    \includegraphics[width=1\textwidth]{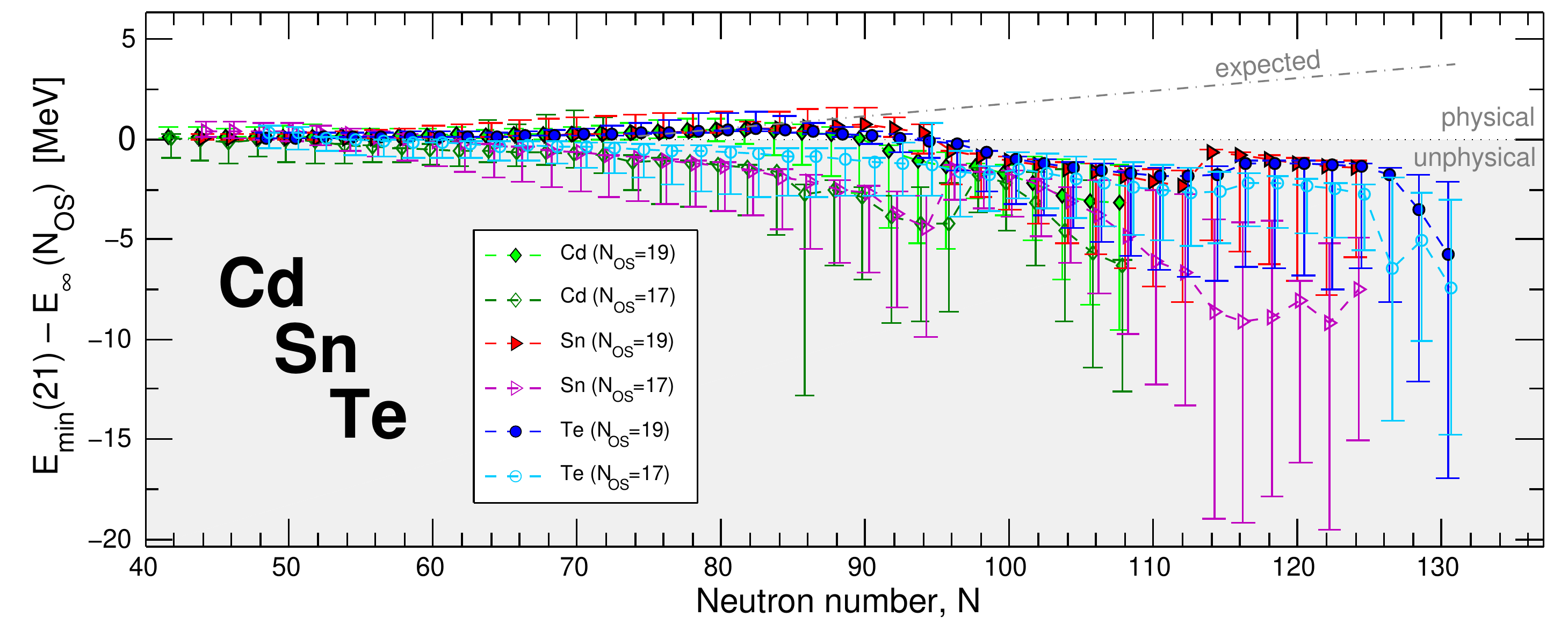}
    \caption{(color online) The plot shows the results of IR extrapolation for nuclei in cadmium, tin, and tellurium isotopic chains as the function of neutron number in form of the differences $E_\mathrm{min}(21)-E_\infty(N_{OS})$  with momentum threshold \hbox{$\Lambda_{thr}=750$ MeV/c}.  Two dimensions of model space is considered, $N_{OS}=17$ (hollow symbols) and $N_{OS}=19$ (filled symbols), separately. The associated errorbars represent the spread in corresponding $E_\infty$ values from the variation of the threshold parameter in the interval 700 MeV/c $<\Lambda_{thr}<900$ MeV/c.}
    \label{Fig12}
  \end{center}
\end{figure*}
%
  \section{Summary and Discussion}
  \label{summary}
%
We have studied the convergence pattern of the HFB energies as a function of the maximum number of SHO shells included in the working basis, $N_{OS}$, as well as a function of the oscillator length, $b$. The calculations were performed with the Gogny D1S EDF. Generally, one has to include a prohibitively large number of $N_{OS}$ in practical calculations to ensure convergence. In order to circumvent this shortcoming, one can opt to use various extrapolation techniques to obtain an estimate of the converged observables. While the ansatz $\Delta E(N_{OS}) \approx \Delta E(N_{OS}-2)/2$, that in central in a purely phenomenological energy correction scheme~\cite{2009PhRvL.102x2501G,2007EPJA...33..237H}, proved generally \textit{not} to hold, we have turned to and studied one of the most promising extrapolation schemes recently proposed, namely, the IR extrapolation~\cite{2015JPhG...42c4032F}.

We have seen that the application of the considered IR extrapolation to a playground case of $^{16}$O seems to work nearly perfect, providing reliable results that are both threshold-independent and consistent with the fully converged reference value. A more serious benchmark, first by application to the nucleus $^{120}$Cd, and then to the whole set of nuclei in the cadmium isotopic chain, revealed, however, some limitations of the proposed scheme. 

Fig.~\ref{Fig12} summarizes the conducted analysis of IR extrapolations scheme for cadmium isotopes, as well as extends the scope of the study to tin and tellurium nuclei. As it is the case with all considered isotopic chains, the most robust extrapolations are obtained for isotopes in the direct vicinity to the stability region. Nevertheless, in this region the extrapolations are least relevant due to the larger degree of convergence of the HFB calculations in comparison to the neutron-rich isotopes. However, as one moves away towards the neutron-drip line, the IR extrapolations fails to yield reliable results. In particular, the discrepancy of the extrapolations from $N_{OS}=17$ and $N_{OS}=19$ values (with $\Lambda_{thr}=750$ MeV/c) reach easily up to 5--8 MeV. Besides that, varying $\Lambda_{thr}$ for neutron--rich nuclei has a much greater impact on the estimated $E_\infty$ values (spanning energies of $10-15$ MeV for $A\sim115$). Finally, we also notice that the IR corrections can no longer even reproduce the most converged HFB 
calculations at hand (i.e. the $E_\mathrm{min}(21)$ values) in the neutron--rich tail of the isotopic chain, which is evident by the negative unphysical estimates for $N\gtrsim96$ on Fig.~\ref{Fig12}. These results have been supported by similar findings for other isotopic chains throughout the nuclear chart~\cite{Arzhanov_master}.

The final conclusion that we can draw, at least in conformity with HFB calculations with Gogny EDF, is that the investigated extrapolation schemes are so far applicable with some reliability only in the regions of well-bound nuclei of light to medium-mass isotopic chains. These restrictions of the proposed IR energy-corrections renders these methods to be of limited applications in astrophysical calculations, as they do not provide reliable estimations at required precision level for heavy nuclei in the vicinity of the neutron-drip line.
\begin{acknowledgments}
  We gratefully thank L. M. Robledo, R. Roth, and A. Rios for fruitful
  discussions. This work was supported by the Ministerio de Econom\'ia
  y Competitividad under Contract No.  FIS-2014-53434, Programa
  Ram\'on y Cajal 2012 No. 11420, the Helmholtz Association through
  the Nuclear Astrophysics Virtual Institute Grant No. VH-VI-417,
  and the BMBF-Verbundforschungsprojekt Grant No. 05P15RDFN1. We also
  acknowledge the support from GSI Darmstadt and LOEWE-CSC Frankfurt
  computing facilities.
\end{acknowledgments}

\bibliography{AA_biblio}

\end{document}